\documentclass[conference]{IEEEtran}
\PassOptionsToPackage{caption=false,font=footnotesize}{subfig}
\usepackage{subfig}  
\IEEEoverridecommandlockouts
\usepackage{cite}
\usepackage{amsmath,amssymb,amsfonts}
\usepackage{textcomp}
\usepackage{xcolor}
\usepackage{array}
\usepackage{scalerel}
\usepackage{multicol}
\usepackage{esdiff}

\usepackage{cases}
\usepackage{lipsum}
\usepackage{balance}
\usepackage{algorithm}
\usepackage{placeins}
\usepackage{graphicx}
\usepackage{textcomp}
\usepackage{algpseudocode}

\usepackage{hyperref}
\usepackage{cleveref}
\usepackage{mathtools}
\usepackage{multirow}

\usepackage{booktabs}
\usepackage[table,dvipsnames]{xcolor} 
\usepackage{tabularx}
\usepackage{threeparttable}
\usepackage{tikz}
\usetikzlibrary{arrows.meta,positioning,fit,shapes.misc}
\usepackage{comment}
\usepackage[none]{hyphenat} % Reduce hyphenation
\usepackage{microtype}      % Enhance justification
\usepackage[none]{hyphenat} % Reduce hyphenation
\usepackage{microtype}      % Enhance justification
\graphicspath{{../PNG_TPEC/}}

\def\BibTeX{{\rm B\kern-.05em{\sc i\kern-.025em b}\kern-.08em
		T\kern-.1667em\lower.7ex\hbox{E}\kern-.125emX}}
\begin{document}
	
	\title{Event-Driven Deep RL Dispatcher for Post-Storm Distribution System Restoration\\
    \thanks{This work was supported by the LSU Institute for Energy Innovation and the LSU Provost’s Fund for Innovation in Research.
    
    \vspace{5pt}
    \text{979-8-3315-4112-5/25/\$31.00 ©2026 IEEE}
    }

		%\thanks{This work was supported by the National Science Foundation under Grant ECCS-1944752 and Grant EECS-2312086. Amin Kargarian is the corresponding author. Conference ID: 63981X \\
        
        %979-8-3315-4112-5/25/\$31.00 ©2025 IEEE}
	}
	
	\author{\IEEEauthorblockN{Farshad Amani}
		\IEEEauthorblockA{\textit{Electrical and Computer Engineering}\\
			\textit{Department} \\
			\textit{Louisiana State University}\\
			Baton Rouge, USA \\
			famani1@lsu.edu}
		\and
		\IEEEauthorblockN{ Faezeh Ardali}
		\IEEEauthorblockA{\textit{Industrial Engineering}\\
			\textit{Department} \\
			\textit{Louisiana State University}\\
			Baton Rouge, USA \\
			fardal1@lsu.edu}
		\and
		\IEEEauthorblockN{ Amin Kargarian}
		\IEEEauthorblockA{\textit{Electrical and Computer Engineering}\\
			\textit{Department} \\
			\textit{Louisiana State University}\\
			Baton Rouge, USA \\
			kargarian@lsu.edu}
	}

	\maketitle
%%%%%%%%%%%%%%%%%%%%%%%%%%%%%%%%%%%%%%%%%%%%%%%%%%%%%%%%%%%%%%%%%%%%%%%%%%%%%%%%%%%%%%%%%%%%%%%%%%%%%%%%%%%%%%%%%%%%%%%%%%%%%%%%%%%%%%%%%%%%%%%%%%%%%%%%%%%%%%%%%%%%%%%%%%%%%%%%%%%%%%%%%%%%%%%%%%%%%%%%%%%%%%%%%%%%%%%%%%%%%%%%%%%%%%%%%%%%%%%%%%%%%%%%%%%%%%%%%%%%%%%%%%%%%%%%%%%%%%%%%%%%%%%%%%%%%%%%%%%%%%%%%%%%%%%%%%%%
\begin{abstract}
Natural hazards such as hurricanes and floods damage power grid equipment, forcing operators to replan restoration repeatedly as new information becomes available. This paper develops a deep reinforcement learning (DRL) dispatcher that serves as a real-time decision engine for crew–to–repair assignments. We model restoration as a sequential, information-revealing process and learn an actor–critic policy over compact features such as component status, travel/repair times, crew availability, and marginal restoration value. A feasibility mask blocks unsafe or inoperable actions, such as power flow limits, switching rules, and crew-time constraints, before they are applied. To provide realistic runtime inputs without relying on heavy solvers, we use lightweight surrogates for wind and flood intensities, fragility-based failure, spatial clustering of damage, access impairments, and progressive ticket arrivals. In simulated hurricane and flood events, the learned policy updates crew decisions in real time as new field reports arrive. Because the runtime logic is lightweight, it improves online performance (energy-not-supplied, critical-load restoration time, and travel distance) compared with mixed-integer programs and standard heuristics. The proposed approach is tested on the IEEE 13- and 123-bus feeders with mixed hurricane/flood scenarios.

\end{abstract}

%%%%%%%%%%%%%%%%%%%%%%%%%%%%%%%%%%%%%%%%%%%%%%%%%%%%%%%%%%%%%%%%%%%%%%%%%%%%%%%%%%%%%%%%%%%%%%%%%%%%%%%%%%%%%%%%%%%%%%%%%%%%%%%%%%%%%%%%%%%%%%%%%%%%%%%%%%%%%%%%%%%%%%%%%%%%%%%%%
\section{Introduction}

\IEEEPARstart{N}{atural} hazards such as hurricanes and floods are among the most disruptive threats to modern power systems. These events often degrade multiple infrastructures at once—electric power, transportation, and communications—and can affect millions of customers, with losses reaching tens to hundreds of billions of dollars \cite{amani2025quantum, amirioun2022prioritizing}. In the United States, tropical cyclones account for the largest share of weather disaster losses since 1980, exceeding \$1.5\,trillion and leading in fatalities. The costliest storms—Katrina (2005), Harvey (2017), Ian (2022), and Maria (2017)—produced widespread, prolonged outages~\cite{noaa_costliest_tc_2025}. The community impacts are significant: extended interruptions delay critical care, disrupt local commerce, and strain public health services~\cite{amani2024seismic}. Accordingly, preparation and resilience planning for such hazards are critical priorities for utilities and policymakers.

The power system preparation process includes three phases. Before the event, utilities harden the system via targeted upgrades, undergrounding, sectionalizing, vegetation management, and adding redundancy. During the event, operators adapt in real time, switching, islanding, coordinating with DERs and microgrids, and adjusting operations as conditions evolve. After the incident, the priority is restoration—coordinating field crews, ensuring safe switching, and mobilizing resources to quickly restore service \cite{amani2025learning}. While all three matter, early restoration decisions are often decisive for reducing energy not supplied (ENS) and limiting human and economic harm \cite{amani2025learning, arif2018optimizing}.

Many mathematical approaches formulate post-event restoration—crew dispatch and switching—as mixed-integer optimization \cite{arjomandi2020modeling}. These models are expressive but NP-hard and slow to solve at operational scale, even with decomposition, rolling horizons, and warm starts \cite{amani2025learning, lotfabadi2025cost}. In practice, utilities also use simple heuristics—priority rules (e.g., critical loads first), nearest-neighbor routing, and MST-based reconnection—plus a few standard metaheuristics (e.g., tabu search, genetic algorithms) and occasional dynamic-programming or network-flow approximations for radial reconfiguration \cite{dolatyabi2025heterogeneous, arthur2025community}. The challenge intensifies as new damage reports arrive and plans must be revised repeatedly.

Because information is revealed gradually, operators need a fast, real-time method that is simple to apply in practice. Machine-learning methods offer practical alternatives \cite{dolatyabi2025deep}. Prior work uses supervised predictors to score candidate repairs based on expected benefits (for example, load restored or customer-minute saved). It updates these scores online as new field reports arrive \cite{eshkaftaki2024resilience}. Authors in \cite{wang2023data} fuse network topology, load, and access signals to build simple priority lists that stay stable even when inputs are noisy. Learning-augmented search pairs fast neural scoring with look-ahead, allowing dispatch plans to refresh quickly as new faults are confirmed \cite{shuai2023post}. Imitation-learning approaches clone expert or solver trajectories, allowing actions such as tie-switching or reactive support to be issued in real-time without solving a large optimization problem. Uncertainty-aware models report a confidence measure with each recommendation, helping crews prioritize inspections and de-risk early moves \cite{oelhaf2025scoping}. Transfer and meta-learning help with cold starts by adapting models to new feeders or event types using knowledge from past storms. Reinforcement learning (RL) is also used—often in multi-agent or graph-based forms—for rapid adaptation as confirmations arrive \cite{goda2025electric, qiu2023hierarchical}.

RL, especially actor–critic methods, optimizes long-horizon objectives directly: the actor selects crew-to-component assignments while the critic stabilizes training, enabling millisecond-to-second replanning as new components are confirmed. These methods trade exact optimality for speed and adaptability—valuable during evolving events \cite{zhang2025deep}. This paper develops a deep RL (DRL) crew-dispatch approach for hurricane/flood restoration that (i) treats restoration as a sequential, information-revealing process over depots, crews, and damaged components; (ii) learns an actor–critic policy that assigns crews in real time using simple, interpretable features (component status, travel/repair time estimates, crew availability, and a marginal restoration value signal); and (iii) includes a fast feasibility screen so unsafe or inoperable actions are masked before execution (power-flow limits, switching rules, crew-time constraints). We deliberately avoid heavy hazard solvers in the control loop. Instead, we generate realistic inputs with lightweight surrogates (wind/flood intensities, fragility, spatial clustering, access impairment, and progressive ticket arrivals) and feed those streams to the dispatcher.

Compared with repeatedly solving large optimization problems, the learned policy (1) offers real-time updates for frequent replans, (2) adapts on the fly to late-arriving field information without re-tuning model structure, (3) scales with data rather than enumerating combinatorial search trees, and (4) aligns with operator goals via reward shaping that prioritizes ENS reduction, critical-load restoration time, and travel efficiency. DRL thus provides a pragmatic path to faster and more resilient restoration during extended storm operations.

The contributions of the paper are listed as follows:
\begin{itemize}
  \item We model crew dispatch for storm/flood restoration as a sequential, partially observable decision process and define compact state/action representations that are compatible with utility data streams.
  \item We implement a model-free actor–critic policy that selects next repairs and crew assignments in real time, with a feasibility mask to maintain safety and operability.
  \item We outline a rollout-based training protocol and evaluate online quality (ENS, time-to-critical-load, travel overhead) against optimization baselines and simple heuristics.
  \item We show how the DRL layer integrates with outage management and switching workflows, providing instant recommendations while allowing periodic solver re-anchoring.
\end{itemize}

%%%%%%%%%%%%%%%%%%%%%%%%%%%%%%%%%%%%%%%%%%%%%%%%%%%%%%%%%%%%%%%%%%%%%%%%%%%%%%%%%%%%%%%%%%%%%%%%%%%%%%%%%%%%%%%%%%%%%%%%%%%%%%%%%%%%%%%%%%%%%%%%%%%%%%%%%%%%%%%%%%%%%%%%%%%%%%%%%%%%
% ===========================
% ===========================
\section{Event Context and Modeling Assumptions}

This section focuses on large weather-driven disruptions that damage distribution networks over wide areas, resulting in new outages that persist for hours to days. This paper uses physics and practice-informed surrogates to (i) generate plausible spatial damage patterns and (ii) stream newly confirmed outages to operations. This keeps the inputs realistic while remaining fast.
%%%%%%%%%%%%%%%%%%%%%%%%%%%%%%%%%%%%%%%%%%%%%%%%%%%%%%%%%%%%%%%%%%%%%%%%%%%%%%%%%%%%%%%%%%%%
\subsection{Hurricane Wind Field and Wind–Induced Damage}

For tropical cyclones \cite{holland2010revised} adopt the symmetric Holland gradient–wind profile to compute site wind speeds from a compact set of event parameters \((\Delta p, R_m, B, \rho_a)\) and a background (translation) term \(V_{\text{bg}}\). The wind at range \(r\) from the storm center is
{\small\begin{equation}
V(r) \;=\;
\sqrt{\left(\frac{B\,\Delta p}{\rho_a}\right)\!\left(\frac{R_m}{r}\right)^{\!B}
\exp\!\Big[-\big(R_m/r\big)^{\!B}\Big]}\;+\;V_{\text{bg}}.
\label{eq:holland}
\end{equation}}
Given asset-level winds \(V_i=V(r_i)\), damage exceedance for a damage state \(ds\) is modeled with a lognormal fragility:
{\small \begin{equation}
\Pr\{\mathrm{damage}_i \ge ds \mid V_i\}
=\Phi\!\left(\frac{\ln\!\left(V_i/V_{50,ds}\right)}{\beta_{ds}}\right),
\label{eq:windfrag}
\end{equation}}
where \(V_{50,ds}\) and \(\beta_{ds}\) are component-class medians and dispersions (e.g., poles, laterals, risers, substation equipment), consistent with FEMA HAZUS hurricane methodology \cite{schneider2006hazus}.

\begin{table}[t]
\centering
\caption{Holland wind-field parameters and typical choices.}
\vspace{-6pt}
\label{tab:holland_params}
\footnotesize
\begin{tabular}{lll}
\hline
\textbf{Symbol} & \textbf{Units} & \textbf{Typical / Source} \\
\hline
\(\Delta p\)      & hPa           & Best-track / reanalysis \\
\(R_m\)           & km            & Best-track / reanalysis \\
\(B\)             & --            & \(1\!-\!3\); \(B \approx 2 - (p_{\rm env}-p_c)/160\) \cite{holland2010revised, vickery2008statistical} \\
\(\rho_a\)        & kg m\(^{-3}\) & \(1.15\) \cite{holland2010revised} \\
\(V_{\text{bg}}\) & m s\(^{-1}\)  & From storm motion/asymmetry \cite{powell2009estimating} \\
\hline
\end{tabular}
\end{table}

%%%%%%%%%%%%%%%%%%%%%%%%%%%%%%%%%%%%%%%%%%%%%%%%%%%%%%%%%%%%%%%%%%%%%%%%%%%%%%%%%%%%%%%%%%%%
\subsection{Flood Depth Field and Flood–Induced Damage}

Following \cite{balbi2023bayesian}, pluvial/fluvial flood depth is represented as a baseline inundation field plus a spatially correlated perturbation:
\begin{equation}
D(x) \;=\; \bar{D}(x;\,\text{event}) \;+\; \varepsilon(x), 
\qquad \varepsilon(x)\sim \mathcal{GP}\!\big(0,\,C_\ell(\|x-x'\|)\big),
\label{eq:flooddepth}
\end{equation}
with an exponential covariance \(C_\ell(h)=\sigma^2 \exp(-h/\ell)\). The variance \(\sigma^2\) controls the magnitude of local departures from the baseline, and the range \(\ell\) governs the spatial scale of correlation (larger \(\ell\) implies broader, smoother variations).

Given the depth at asset location \(x_i\), \(D_i=D(x_i)\), damage exceedance for state \(ds\) is modeled with a lognormal fragility:
{\small
\begin{equation}
\Pr\{\text{damage}_i \ge ds \mid D_i\}
=\Phi\!\left(\frac{\ln\!\left(D_i/D_{50,ds}\right)}{\beta_{ds}}\right),
\label{eq:floodfrag}
\end{equation}}
where \(D_{50,ds}\) and \(\beta_{ds}\) denote, respectively, the median depth and dispersion for component class \(i\). Parameters may incorporate local elevation or engineered protection by adjusting the baseline \(\bar{D}\) and/or the medians \(D_{50,ds}\).

%%%%%%%%%%%%%%%%%%%%%%%%%%%%%%%%%%%%%%%%%%%%%%%%%%%%%%%%%%%%%%%%%%%%%%%%%%%%%%%%%%%%%%%%%%%%

\subsection{Correlated Multi–Hazard Sampling and Access Impairment}

Nearby assets tend to fail together. A Gaussian copula induces this spatial clustering while preserving each site’s marginal failure probability from wind and flood \cite{zeng2020modelling}.

To capture clustering, site-level Bernoulli damage indicators are coupled with a Gaussian copula using covariance \(C_\ell\). Let \(U_i=\Phi(Z_i)\) with \((Z_1,\ldots,Z_n)\sim \mathcal{N}(0,\Sigma)\) and \(\Sigma_{ij}=C_\ell(\|x_i-x_j\|)\). Set
{\small
\begin{equation}
z_i(0) \;=\; \mathbf{1}\{\,U_i < p_i\,\}, 
\qquad p_i = 1 - \prod_{h\in\{\text{wind},\,\text{flood}\}}\big(1-p_i^{(h)}\big),
\label{eq:copula}
\end{equation}}
where \(p_i^{(\text{wind})}\) and \(p_i^{(\text{flood})}\) come from \eqref{eq:windfrag}–\eqref{eq:floodfrag}. 

The combined probability \(p_i\) is the chance that at least one hazard causes damage; the copula then correlates the draws so nearby \(z_i(0)\) are more likely to co-occur.

\medskip
Roads can also be impaired. Each segment \(e\) is marked unavailable with probability \(q_e\) (from wind/flood thresholds). Crews route on the remaining open network, with travel times inflated to reflect debris, detours, and congestion \cite{amini2023probabilistic}:
\vspace{-4pt}
\begin{equation}
\tau_{k\to i}(t) \;=\; \rho_t \,\frac{d_{k,i}^{\text{open}}(t)}{v_k}, 
\qquad \rho_t \in [\underline{\rho},\overline{\rho}],
\label{eq:travel}
\end{equation}
where \(d_{k,i}^{\text{open}}(t)\) is the shortest-path distance on the open-road graph. The factor \(\rho_t\) is a simple knob for network-wide slowdowns; larger \(\rho_t\) captures heavier congestion or obstructions.

%%%%%%%%%%%%%%%%%%%%%%%%%%%%%%%%%%%%%%%%%%%%%%%%%%%%%%%%%%%%%%%%%%%%%%%%%%%%%%%%%%%%%%%%%%%%
\subsection{Progressive Outage Discovery and Repair Times}

Following a major event, the set of known outages evolves over hours to days as field inspections proceed, customers regain connectivity, and telemetry is restored. To represent this reporting pipeline, confirmed outage counts are modeled by a nonhomogeneous Poisson process with time-varying rate $\lambda(t)$:
{\small \begin{equation}
\label{eq:nhpp}
\begin{aligned}
N(t) &\sim \operatorname{Poisson}\!\left(\int_{0}^{t}\lambda(u)\,du\right),\\
B_t &\sim \operatorname{Poisson}\!\big(\lambda(t)\,\Delta t\big),
\end{aligned}
\end{equation}}
where $\lambda(t)$ governs the pace of discovery (larger values produce more reports per unit time). To capture strictly positive, right–skewed durations typical of field work, repair times are modeled as lognormal by component class $i$:
\begin{equation}
\log \tau_i^{\mathrm{rep}} \sim \mathcal{N}(\mu_i,\sigma_i^2).
\label{eq:repair}
\end{equation}
This choice yields realistic long-tail behavior without imposing arbitrary hard bounds, and it is straightforward to calibrate via maximum likelihood (or moment matching) on historical work orders. 

%%%%%%%%%%%%%%%%%%%%%%%%%%%%%%%%%%%%%%%%%%%%%%%%%%%%%%%%%%%%%%%%%%%%%%%%%%%%%%%%%%%%%%%%%%%%
\subsection{Stochastic Scenario Sampling for Dispatch}

We turn hazard summaries into dispatch-ready samples that (i) convert wind/flood intensity into damage likelihoods, (ii) respect spatial clustering and delayed discovery, and (iii) embed access and repair-time frictions. These stochastic sequences feed the event-driven dispatcher.

Given $\{\Delta p,R_m,B,V_{\text{bg}}\}$ for hurricanes or a flood baseline map, asset classes, and correlation length $\ell$: 
\begin{enumerate}
  \item Compute site winds $V_i$ via \eqref{eq:holland} and/or depths $D_i$ via \eqref{eq:flooddepth}.
  \item Map intensities to damage probabilities with \eqref{eq:windfrag}–\eqref{eq:floodfrag} and combine $p_i$ across hazards.
  \item Draw correlated outages $z_i(0)$ via \eqref{eq:copula}; set $\mathcal{D}_0=\{i:z_i(0)=1\}$.
  \item Evolve discoveries with \eqref{eq:nhpp} to obtain $\Delta\mathcal{D}_t$ and update $\mathcal{D}_{t+1}=\mathcal{D}_t\cup\Delta\mathcal{D}_t$.
  \item For routing/scheduling, compute travel times $\tau_{k\to i}(t)$ via \eqref{eq:travel} and sample repair times $\tau_i^{\mathrm{rep}}$ via \eqref{eq:repair}.
\end{enumerate}

Steps 1–4 produce spatially clustered, time-revealed damage consistent with the hazard; Step 5 encodes operational frictions (access/congestion) and stochastic work durations. Fixing the random seed enables repeatable experiments; varying $\ell$, $\lambda(t)$, or class parameters supports sensitivity analysis.

%%%%%%%%%%%%%%%%%%%%%%%%%%%%%%%%%%%%%%%%%%%%%%%%%%%%%%%%%%%%%%%%%%%%%%%%%%%%%%%%%%%%%%%%%%%%%%%%%%%%%%%%%%%%%%%%%%%%%%%%%%%%%%%%%%%%%%%%%%%%%%%%%%%%%%%%%%%%%%%%%%%%%%%%%%%%%%%%%%%%%%%%%%%%%%%%%%%%%%%%%%%%%%%%%%%%%%%%%%%%%%%%%%%%%%%%%%%%%%%%%%%%%%%%%%%%%%%%%%%%%%%%%%%%%%%%%%%%
% ===========================
% SECTION 2: DRL dispatch 
% ===========================
\section{DRL for Crew Dispatch After Events}
We model post-event restoration as a sequential decision problem in which multiple repair crews must be routed and rerouted as new information arrives. Immediately after the incident, an initial set of damaged components is known; additional damage reports appear over time (e.g., aftershocks, late inspections). The dispatcher must react quickly, reprioritizing work to minimize service interruptions while maintaining safety and operability.

\subsection{Dynamic MDP and Reward}
Decision-making is cast as a partially observable MDP $\mathcal{M}=(\mathcal{S},\mathcal{A},P,R,\gamma)$. At decision epoch $t$, the observed state $s_t$ aggregates component, crew, and global features,
\vspace{-4pt}
\begin{equation}
  s_t \;=\; \big[\, x_t^{\mathrm{comp}},\; x_t^{\mathrm{crew}},\; x_t^{\mathrm{global}} \,\big],
  \label{eq:state}
\end{equation}
%\vspace{-4pt}
where $x_t^{\mathrm{comp}}$ includes damage flags $z_i(t)$, estimated repair duration $\tau_i^{\mathrm{rep}}$, access-aware travel time $\tau_{k\to i}(t)$, and marginal restoration value $v_i(t)$; $x_t^{\mathrm{crew}}$ includes crew locations, availability, skills, and remaining shift; $x_t^{\mathrm{global}}$ tracks time, remaining unserved demand, and critical-load indicators.
Available crews receive assignments via the joint action
\vspace{-4pt}
{\small\begin{equation}
  a_t \;=\; \big\{\, a_t^{(k)} \in \mathcal{C}\cup\{\mathrm{hold},\mathrm{return}\} \,\big\}_{k\in\mathcal{K}} .
  \label{eq:action}
\end{equation}}
Safety and operability are enforced by a feasibility gate that masks illegal choices before sampling or greedy selection:
{\small \begin{equation}
  m_t(a) \;=\; \mathbf{1}\!\left\{\, g(s_t,a)\le 0 \;\wedge\; h(s_t,a)=0 \right\},
  \label{eq:mask}
\end{equation}}
where $g$ collects inequality constraints (e.g., voltage and branch limits, crew-time rules) and $h$ encodes switching/topology logic. Masked actions receive $-\infty$ logit before the softmax.

We use a dense reward aligned with restoration goals,
{\small \begin{align}
\small
  r_t \;=\;& -\alpha\,\mathrm{ENS}_t \;-\; \beta\,\mathrm{Travel}_t \;-\; \gamma\,\mathrm{Idle}_t \nonumber\\
           & -\eta\,\mathrm{Viol}_t \;+\; \kappa\,\mathrm{CritRestored}_t.
\label{eq:reward}
\end{align}}
and maximize the discounted return
\vspace{-4pt}
{\small\begin{equation}
  \max_{\pi_\theta}\; J(\theta) \;=\; \mathbb{E}_{\pi_\theta}\!\Big[\sum_{t=0}^{T}\gamma^t r_t\Big].
  \label{eq:objective}
\end{equation}}
\vspace{-6pt}
\subsection{Policy, Value, and Memory}
We adopt an actor--critic architecture. The actor $\pi_\theta(a\mid s_t)$ outputs crew-wise logits with feasibility masks applied before the softmax; the critic $V_\psi(s_t)$ provides a baseline for variance reduction. Because observations are partial and the damage set evolves, we maintain a compact memory state,
\begin{align}
\small
  h_t &= \mathrm{RNN}(h_{t-1},\, s_t), \nonumber\\
  \pi_\theta(a\mid s_t) &= \pi_\theta(a\mid h_t), \nonumber\\
  V_\psi(s_t) &= V_\psi(h_t).
  \label{eq:rnn}
\end{align}

Policy updates use advantages (with $n$-step/GAE variants),
\begin{equation}
  A_t \;=\; r_t + \gamma V_\psi(s_{t+1}) - V_\psi(s_t).
  \label{eq:adv}
\end{equation}
\vspace{-6pt}
\subsection{Event-Driven Interaction}
The environment is event-driven: the clock advances to the next salient event (ticket arrival, travel completion, repair completion, duty change). At each event, the state is updated, and the agent may replan if the conditions warrant it. This avoids unnecessary decisions when nothing material has changed, aligning with field operations. Algorithm 1 is the event-driven simulator and online dispatcher that steps through outage events in time to choose safe crew dispatch and switching actions.

\begin{algorithm}[t]
\small
\caption{Event-driven environment and online dispatch}
\label{alg:env}
\begin{algorithmic}[1]
\State Initialize network $\mathcal{K}_0$, crews $\mathcal{C}_0$, initial damages $\mathcal{D}_0$, and clock $\tau_0$.
\While{time horizon not reached \textbf{and} unserved load remains}
  \State Advance the clock to the next event (new ticket, travel end, repair end, duty change) and update ENS.
  \If{new tickets arrive}
    \State Add them to $\mathcal{D}$.
  \EndIf
  \If{a crew arrives on site}
    \State Start or continue the repair if it is feasible.
  \EndIf
  \If{a repair finishes}
    \State Update $\mathcal{K}$ and re-energize the network safely.
  \EndIf
  \State Form $s_t = \{\mathcal{D},\mathcal{C},\mathcal{K},\tau\}$ and update $h_t$ via~\eqref{eq:rnn}.
  \If{a replan trigger fires (event-based or periodic)}
     \State Build feasibility masks per~\eqref{eq:mask}.
     \State Select $a_t$ from $\pi_\theta(\cdot\mid h_t)$ (masked sampling or greedy).
     \State Commit assignments; send crews and apply safe switching.
  \EndIf
\EndWhile
\end{algorithmic}
\end{algorithm}
\vspace{-8pt}
\subsection{Learning: PPO with Masked Actions}
We train with proximal policy optimization (PPO). For each rollout, the agent interacts with the event-driven simulator to collect trajectories $\{(s_t,a_t,r_t,s_{t+1},\text{mask}_t,\log\pi_\theta(a_t\mid s_t))\}$.
Advantages are computed with the critic (GAE). The actor is optimized with the clipped objective; the value network regresses returns; and an entropy bonus promotes exploration. Feasibility masks are deterministic and enforced at the logit level, so unsafe or inoperable actions are never sampled. Algorithm 2 shows the PPO training with feasibility masks.

\begin{algorithm}[t]
\small
\caption{PPO training for dynamic crew dispatch}
\label{alg:ppo}
\begin{algorithmic}[1]
\State Initialize actor $\theta$, critic $\psi$, clip $\epsilon$, coefficients $(c_v,c_e)$, discount $\gamma$, and GAE parameter $\lambda$.
\For{epoch $=1,\dots,E$}
  \State $\mathcal{B}\!\leftarrow\!\emptyset$.
  \For{each rollout}
     \State Reset environment; $s\!\leftarrow\! s_0$, $h\!\leftarrow\! h_0$.
     \While{episode not done}
        \State Build masks from $s$; sample $a\!\sim\!\pi_\theta(\cdot\mid h)$ with masks; step to $s',r$.
        \State Store $(s,a,r,s',\text{mask},\log\pi_\theta(a\mid s))$ in $\mathcal{B}$; set $s\!\leftarrow\! s'$.
     \EndWhile
  \EndFor
  \State Compute returns $\hat R$ and advantages $\hat A$ using $V_\psi$.
  \For{update iter $=1,\dots,U$}
     \State Recompute masked log-probs and importance ratios; optimize clipped policy, value loss, and entropy bonus.
     \State Clip gradients; update $(\theta,\psi)$.
  \EndFor
  \State Evaluate on held-out dynamic scenarios (greedy or low-temperature selection).
\EndFor
\end{algorithmic}
\end{algorithm}
%%%%%%%%%%%%%%%%%%%%%%%%%%%%%%%%%%%%%%%%%%%%%%%%%
\vspace{-8pt}
\subsection{Deployment}

\begin{figure}[!t]
\centering
\small
\resizebox{0.98\columnwidth}{!}{%
\begin{tikzpicture}[
  node distance=2.8mm and 5mm,
  box/.style={draw, rounded corners=2pt, inner sep=2pt, align=center},
  >=Latex
]

% Left: runtime sources
\node[box] (src1) {Wind/Flood\\surrogates};
\node[box, below=of src1] (src2) {Access\\roads,\ $\rho_t$};
\node[box, below=of src2] (src3) {Discovery\\$\lambda(t)$};
\node[box, below=of src3] (src4) {Repair priors\\by class};

% Streams
\node[box, right=9mm of src2] (streams) {Streams:\\
$\mathcal D_0,\ \Delta\mathcal D_t,\ \tau_{k\to i}(t),\ \tau_i^{\rm rep}$};

% Runtime loop
\node[box, right=10mm of streams] (env) {Event-driven env.\\(next arrival/travel/repair)};
\node[box, above=of env] (state) {State $s_t$};
\node[box, right=of state] (mask) {Feasibility mask};
\node[box, right=of mask] (actor) {Policy $\pi_\theta$\\(greedy/low-$T$)};
\node[box, below=of actor] (apply) {Apply action\\dispatch \& switching};

% Outcomes
\node[box, right=9mm of apply] (metrics) {Outcomes:\\ ENS,\ crit.\ time,\ travel,\ decision runtime};

% Arrows (sources -> streams)
\draw[->] (src1) -- (streams);
\draw[->] (src2) -- (streams);
\draw[->] (src3) -- (streams);
\draw[->] (src4) -- (streams);

% Arrows (runtime flow)
\draw[->] (streams) -- (env);
\draw[->] (env) -- (state);
\draw[->] (state) -- (mask);
\draw[->] (mask) -- (actor);
\draw[->] (actor) -- (apply);
\draw[->] (apply) -- (env);
\draw[->] (apply) -- (metrics);

\end{tikzpicture}%
}
\vspace{-6pt}
\caption{Deployment dataflow. Surrogate wind and flood models generate runtime streams, and an event-driven loop builds the state $s_t$, masks infeasible actions, selects crew and switching assignments, and applies them with a low decision runtime.}
\label{fig:overview}
\end{figure}
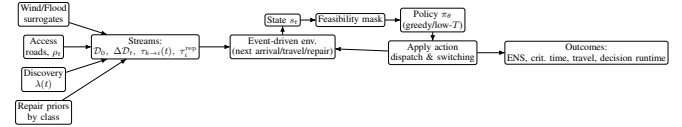

Figure~\ref{fig:overview} summarizes the runtime dataflow. Event surrogates (wind/flood, access, discovery, repair priors) generate streams of tickets, travel times, and repair durations that drive an event-based environment advancing the clock to the next arrival, travel/repair completion, or duty change. At each event, we form the state $s_t$, apply a feasibility mask to remove unsafe/inoperable choices, and select actions via a greedy or low-temperature policy. Assignments are committed immediately (dispatch, safe switching) and the loop repeats as new information arrives, keeping decisions aligned with \eqref{eq:reward} while remaining safe and real-time deployable.

\section{Case Study}
\label{sec:case}

We test the method on storm scenarios with evolving information, impaired access, and class-based repair times. We aim to measure online quality and decision runtime under frequent replanning and to compare a learned policy with an optimization baseline and fast heuristics.
\vspace{-8pt}
\subsection{Experimental Setup}
\label{sec:setup}

We use two standard distribution feeders to maintain consistency and reproducibility: the IEEE 13-bus system with three depots, and the IEEE 123-bus system with six depots. Critical loads (clinics, water pumps, shelters) are tagged in both systems. Crews start from assigned depots, travel on the open-road graph at class-dependent speeds, work 12-hour shifts with mandatory breaks, and must respect switching interlocks. Before any action, a feasibility screen masks choices that would violate power-flow or crew-time rules.

Scenarios are generated using the surrogates in Section II. For hurricanes, we sample Holland-field parameters \((\Delta p,R_m,B,V_{\text{bg}})\) to cover Cat~2--4 landfalls with typical track speeds and crossing angles. For floods, we take a baseline depth map and add a spatially correlated perturbation with range \(\ell\) and variance \(\sigma^2\). Wind and depth maps are converted to class-based damage probabilities using the same lognormal fragility model. Multi-hazard coupling uses the union rule with a Gaussian copula to induce spatial clustering. Access is represented by per-segment impairment and a time-varying inflation \(\rho_t\) that captures detours and congestion. Confirmed tickets arrive according to a nonhomogeneous Poisson rate \(\lambda(t)\) with a daylight peak and a post-eye surge. On-site repair times follow bounded or lognormal priors by class with truncation to shift length.

We scale crew counts with system size: for the 13-bus feeder, \(K\in\{3,6,9\}\); for the 123-bus feeder, \(K\in\{6,12,18\}\). Optional switching crews are included to keep the network radial.

Training uses PPO with an RNN memory (hidden size 128), event-driven stepping (the clock jumps to arrivals, travel ends, repair completions, or duty changes), a clipped objective, GAE, and an entropy bonus. Training and test seeds are disjoint. At evaluation time, the policy runs greedy or at low temperature without exploration noise. We compare against: (i) a rolling MILP with a two-hour look-ahead and a hard per-replan time cap, (ii) two fast heuristics (marginal value-per-time; a travel-aware variant), and (iii) an imitation policy trained on solver trajectories where available. All methods use the same feasibility screen and access model.

We report ENS, the 95th percentile time to first restoration of critical loads, total travel distance, decision runtime per replan, and replan count. Safety and operability violations are required to be zero due to the masking effect.

\subsection{DRL Policy Performance}
\label{sec:results}

\begin{figure}[t]
  \centering
  \begin{tabular}{@{}cc@{}}
    \includegraphics[width=0.5\linewidth]{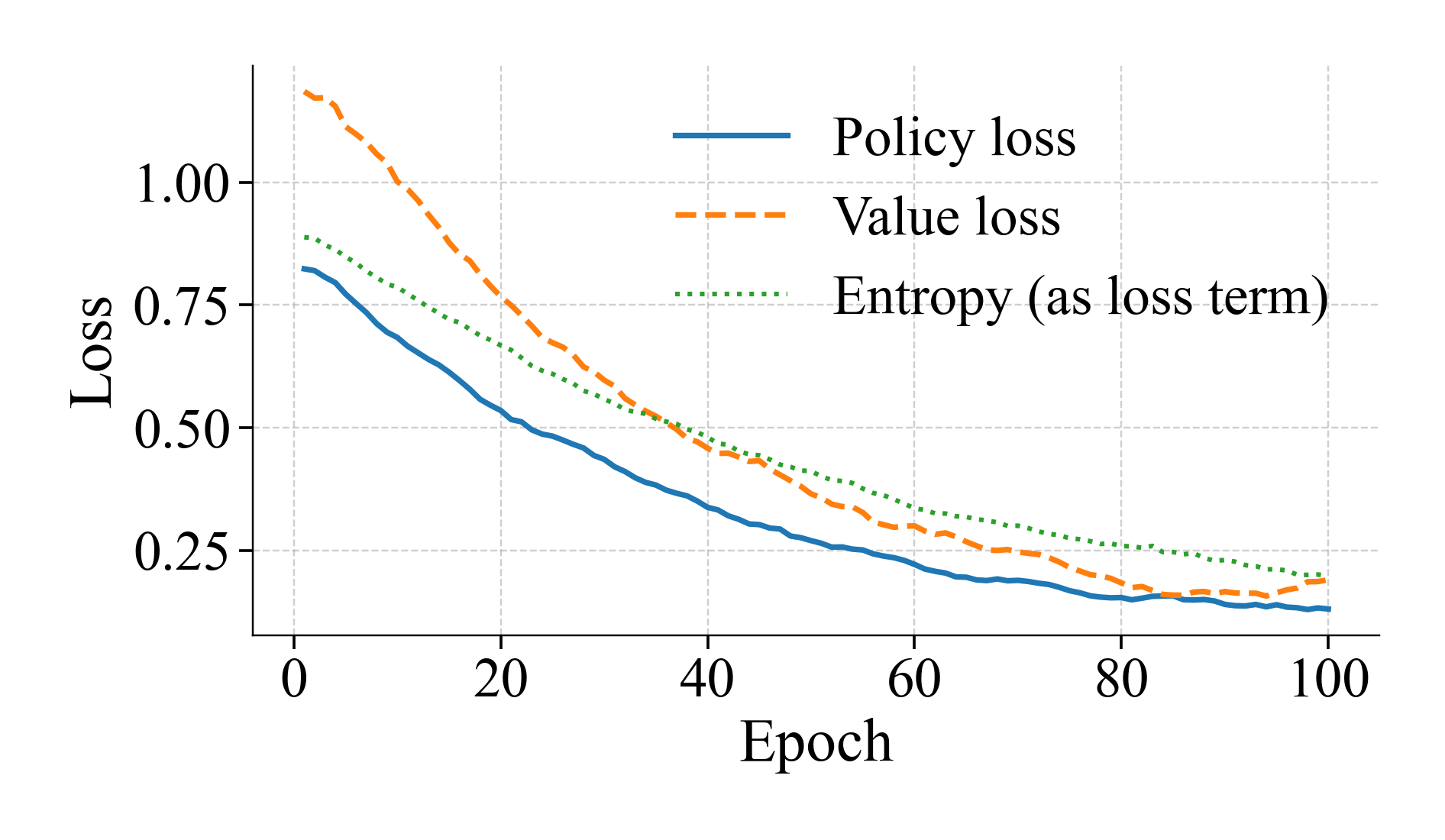} &
    \includegraphics[width=0.5\linewidth]{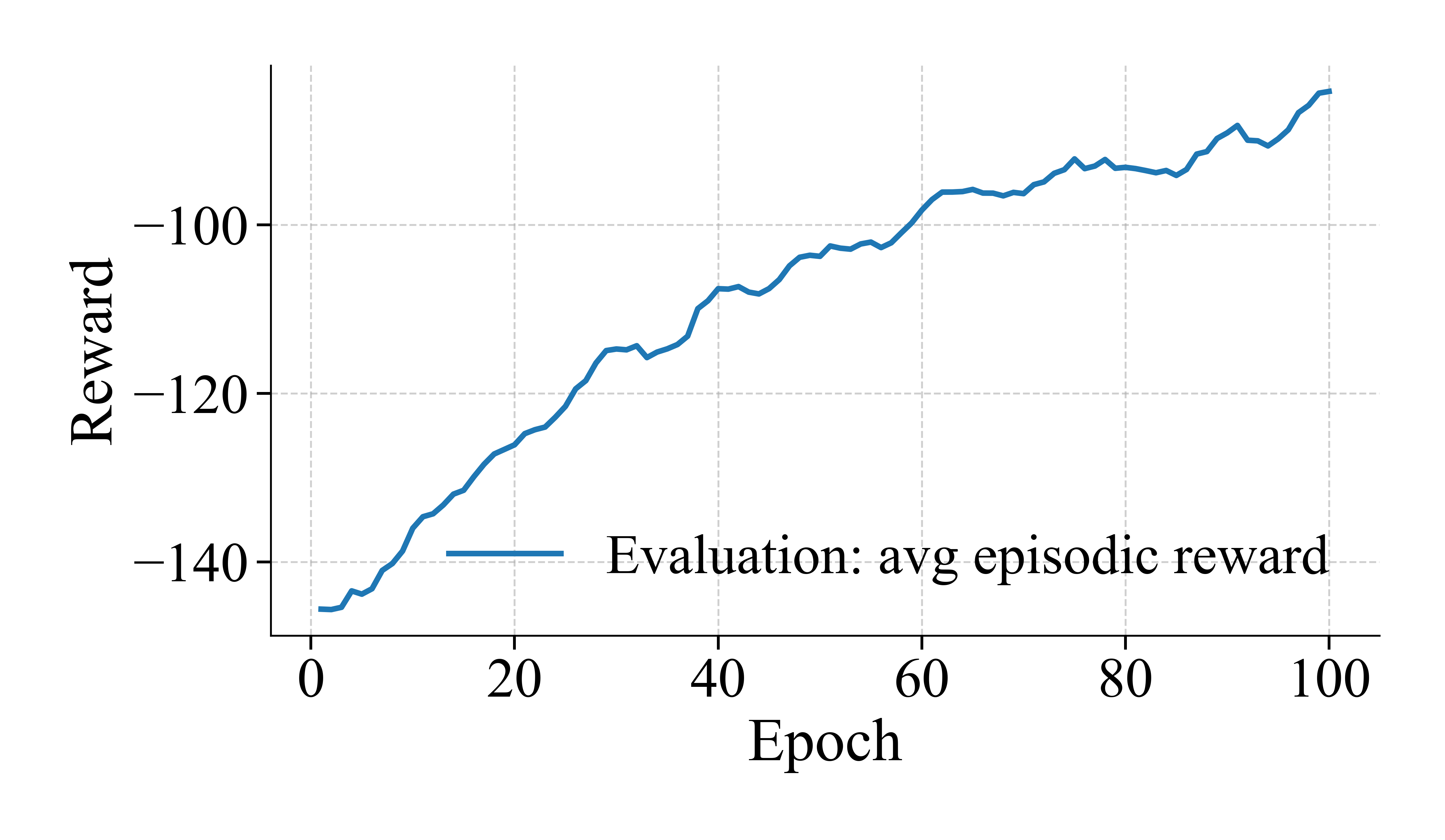}
  \end{tabular}
  \vspace{-8pt}
  \caption{(a) RL training losses over epochs and (b) average episodic reward over epochs.}
  \label{fig:rl_loss_eval}
\end{figure}

The learning curves in Fig.~\ref{fig:rl_loss_eval}(a) show policy and value losses that decay smoothly; entropy (as a loss term) drops as the policy sharpens. Evaluation reward in Fig.~\ref{fig:rl_loss_eval}(b) rises quickly over the first 30--40 epochs, then improves more slowly. We select the best validation-reward checkpoint for test runs.
\begin{comment}
\begin{figure}[t]
  \centering
  \includegraphics[width=0.8\linewidth]{dvrp_rl_training_losses.png}
  \caption{RL training losses over epochs.}
  \label{fig:rl-losses}
\end{figure}

\begin{figure}[t]
  \centering
  \includegraphics[width=0.8\linewidth]{dvrp_rl_eval_reward.png}
  \caption{Evaluation: average episodic reward over epochs (higher is better).}
  \label{fig:rl-eval}
\end{figure}
\end{comment}

Table~\ref{tab:main-iqr} summarizes median performance with interquartile ranges across 300 mixed test scenarios (hurricanes, floods, combined, and a shifted set outside the training range). DRL reduces ENS and critical-load delay compared to fast heuristics while maintaining low decision runtime. Rolling MILP achieves slightly lower travel distance but misses the tight runtime budget by orders of magnitude when replanning frequently.

\begin{table}[t]
\centering
\caption{Median [IQR] over 300 scenarios. Values shown as median [25th--75th pct].}
\label{tab:main-iqr}
\vspace{-8pt}
\setlength{\tabcolsep}{2.5pt}
\renewcommand{\arraystretch}{1.05}
\scriptsize
\begin{tabular}{lcccc}
\toprule
Method & ENS (MWh) & Crit.\ t95 (min) & Travel (km) & Runtime (ms) \\
\midrule
DRL (ours)     & 28\,[22--37]  & 115\,[95--150]   & 410\,[360--470] & 8\,[6--12] \\
MILP (rolling) & 30\,[24--36]  & 120\,[100--155]  & 395\,[350--460] & 30k\,[15k--60k] \\
Greedy-value   & 44\,[35--56]  & 175\,[150--220]  & 520\,[470--590] & 2\,[1--3] \\
Travel-aware   & 39\,[31--50]  & 160\,[135--205]  & 450\,[400--510] & 3\,[2--5] \\
Imitation      & 33\,[26--42]  & 130\,[110--170]  & 430\,[380--495] & 2\,[1--3] \\
\bottomrule
\end{tabular}
\end{table}

DRL is best or tied on ENS and critical-load time under the stated compute limits. MILP manages slightly shorter routes because it optimizes routing explicitly, but its higher runtime makes frequent replanning impractical. Greedy and travel-aware heuristics are fast yet yield higher ENS and slower critical-load restoration. Imitation sits between DRL and heuristics and is more sensitive to distribution shift.

%%%%%%%%%%%%%%%%%%%%%%%%%%%%%%%%%%%%%%%%%%%%%%%%%%%%%%%%%%%%%%%%%%%%%%%%%%%%%%%%%%%%%%%%%%%%
\subsection{System-Level Breakdown}
\label{sec:per-feeder}

To make the picture concrete, we break out the results for the IEEE 13-bus and 123-bus systems. The smaller 13-bus feeder naturally yields lower absolute ENS and travel; the 123-bus feeder stresses routing complexity and timing. All methods share the same feasibility mask and access model; MILP uses the same look-ahead with per-replan time caps.

\begin{table}[t]
\centering
\caption{Per-feeder results (IEEE 13-bus): median [25th--75th pct].}
\vspace{-8pt}
\label{tab:results-13-iqr}
\setlength{\tabcolsep}{2.8pt}
\renewcommand{\arraystretch}{1.05}
\scriptsize
\begin{tabular}{lcccc}
\toprule
Method & ENS (MWh) & Crit.\ t95 (min) & Travel (km) & Runtime (ms) \\
\midrule
DRL (ours)     & 18\,[14--24]  & 85\,[70--110]   & 210\,[180--250] & 6\,[5--9] \\
MILP (rolling) & 19\,[15--24]  & 90\,[75--115]   & 200\,[170--240] & 12k\,[6k--25k] \\
Greedy-value   & 32\,[26--40]  & 135\,[115--170] & 300\,[260--360] & 2\,[1--3] \\
Travel-aware   & 29\,[23--36]  & 125\,[105--160] & 270\,[230--320] & 2\,[1--3] \\
Imitation      & 24\,[19--31]  & 100\,[85--130]  & 260\,[220--310] & 2\,[1--3] \\
\bottomrule
\end{tabular}
\end{table}

\begin{table}[t]
\centering
\caption{Per-feeder results (IEEE 123-bus): median [25th--75th pct].}
\vspace{-8pt}
\label{tab:results-123-iqr}
\setlength{\tabcolsep}{2.8pt}
\renewcommand{\arraystretch}{1.05}
\scriptsize
\begin{tabular}{lcccc}
\toprule
Method & ENS (MWh) & Crit.\ t95 (min) & Travel (km) & Runtime (ms) \\
\midrule
DRL (ours)     & 36\,[29--46]  & 145\,[120--185] & 580\,[520--650] & 10\,[7--14] \\
MILP (rolling) & 38\,[31--46]  & 150\,[125--190] & 560\,[500--630] & 45k\,[25k--80k] \\
Greedy-value   & 58\,[48--72]  & 210\,[180--260] & 720\,[650--820] & 2\,[1--3] \\
Travel-aware   & 51\,[42--64]  & 190\,[165--235] & 630\,[560--710] & 3\,[2--5] \\
Imitation      & 41\,[33--53]  & 165\,[140--210] & 610\,[540--690] & 2\,[1--3] \\
\bottomrule
\end{tabular}
\end{table}

\paragraph{IEEE 13-bus}
DRL delivers the lowest ENS and fastest critical-load restoration under tight compute. Relative to the greedy heuristic, ENS drops from 32 to 18~MWh (\(\approx 44\%\) lower) and Crit.\ \(t_{95}\) falls from 135 to 85~min (\(\approx 37\%\) faster). Against the travel-aware heuristic, ENS improves by \(\approx 38\%\) and Crit.\ \(t_{95}\) by \(\approx 32\%\). MILP edges out DRL on travel (200 vs.\ 210~km median) but takes 12\,000~ms median runtime per replan, versus 6~ms for DRL---roughly a \(2{,}000\times\) gap. IQRs are tight for both DRL and MILP, indicating stable behavior across scenarios.

\paragraph{IEEE 123-bus}
As the problem scales, DRL retains its ENS and critical-load-time advantage while staying in the single- to low double-digit millisecond range. ENS drops from 58 to 36~MWh versus greedy (\(\approx 38\%\) lower) and from 51 to 36~MWh versus the travel-aware variant (\(\approx 29\%\) lower). Crit.\ \(t_{95}\) improves by \(\approx 31\%\) (210 to 145~min) versus greedy and by \(\approx 23\%\) versus travel-aware. MILP yields slightly shorter routes (560 vs.\ 580~km) but needs 45{,}000~ms per replan, making high-frequency replanning impractical. On 123-bus, heuristic IQRs widen, indicating less consistent performance under access variability and clustered discoveries.

\paragraph{Cross-feeder trends}
(i) DRL consistently matches or beats MILP on ENS and critical-load timing while keeping runtime in the few-millisecond range; this is what enables frequent replans when tickets arrive, crews finish, or access improves. (ii) MILP’s shorter travel confirms that explicit route optimization helps with distance, but the runtime cost dominates when the environment changes quickly. (iii) Smaller IQRs for DRL on key outcomes point to smoother behavior under clustered damage and time-varying access; heuristics fluctuate more, especially on the 123-bus system. (iv) All methods respect the same safety/operability masks; within those rules, DRL provides the best quality-runtime balance for online dispatch. Examining each feeder separately does not alter the conclusions. DRL still gives the best overall balance: travel close to MILP but with much faster response, especially on the larger feeder where information changes more quickly.

%%%%%%%%%%%%%%%%%%%%%%%%%%%%%%%%%%%%%%%%%%%%%%%%%%%%%%%%%%%%%%%%%%%%%%%%%%%%%%%%%%%%%%%%%%%%%%%%%%%%%%%%%%%%%%%%%%%%%%%%%%%%%%%%%%%%%%%%%%%%%%%%%%%%%%%%%%%%%%%%%%%%%%%%%%%%%%%%%%%%
\section{Conclusion}

This work introduced a practical event-driven RL framework for post-storm crew dispatch. It uses simple surrogates for hurricane and flood impacts and provides a real-time policy that is easy to extend to new damage components. The surrogates generate realistic streams of tickets, travel times, and repair durations without heavy hazard solvers. The DRL layer reacts at each event and makes safe assignments in milliseconds.

In mixed hurricane and flood scenarios on the IEEE 13-bus and 123-bus systems, the learned policy reduced ENS and shortened critical-load restoration times compared with fast heuristics. The proposed approach achieves near-optimal performance close to a rolling MILP, and the feasibility mask keeps the resulting dispatch plans practical and valid. Because it can be solved in real time and is easy to extend, it can outperform MILP in practice by reducing ENS and speeding up overall restoration.

Since the proposed approach is modular, wind and flood fields, access models, and repair-time priors can be replaced or refined without changing the dispatcher interface, and it fits naturally into existing outage management and switching workflows. It provides instant recommendations in real time, while still allowing periodic re-anchoring with optimization tools. DRL offers a practical way to achieve faster and more resilient restoration during extended storm operations. 

Future work will extend the action space to richer switching plans and multi-crew coordination, integrate faster network checks, and add risk-aware training to handle uncertainty in damage discovery and access. We also plan to test transfer across feeders and utilities and to combine the RL loop with periodic optimization for hybrid operation.

%%%%%%%%%%%%%%%%%%%%%%%%%%%%%%%%%%%%%%%%%%%%%%%%%%%%%%%%%%%%%%%%%%%%%%%%%%%%%%%%%%%%%%%%%%%%%%%%%%%%%%%%%%%%%%%%%%%%%%%%%%%%%%%%%%%%%%%%%%%%%%%%%%%%%%%%%%%%%%%%%%%%%%%%%%%%%%%%%%%%%%%%%%%%%%%%%%%%%%%%%%%%%%%%%%%%

\bibliographystyle{IEEEtran}
\bibliography{TPEC_references}

\end{document}